\documentclass[12pt, onecolumn]{IEEEtran}
\usepackage{amsmath}
\usepackage{amssymb}
\usepackage{mathrsfs}
\usepackage{cite}
\usepackage{epsfig}
\usepackage{epsf}
\usepackage{theorem}
\usepackage{graphics}
\usepackage[active]{srcltx}

\renewcommand{\baselinestretch}{1.85}

\newtheorem{defi}{Definition}

\begin{document}

\title{On the Energy Efficiency of LT Codes in\\
Proactive Wireless Sensor Networks
\thanks{$^{\dag \dag}$The second author completed his Ph.D. in ECE at University of Toronto.}}

\author{\small \textbf{Jamshid Abouei$^\dag$, \emph{Member, IEEE,} J. David
Brown$^{\dag \dag}$, Konstantinos N. Plataniotis$^\dag$, \emph{Senior Member, IEEE} and \\
Subbarayan Pasupathy$^\dag$, \emph{Life Fellow, IEEE}} \\
$^\dag$ \small The Edward S. Rogers Sr. Dept. of Electrical and Computer Engineering, \\
University of Toronto, Toronto, Canada, Emails: \{abouei, kostas, pas\}@comm.utoronto.ca\\
$^{\dag \dag}$ Ottawa, Canada, Tel: 613-236-1051, Email:
david\_jw\_brown@yahoo.com}

\maketitle

\markboth{\small Submitted to IEEE Transactions on Wireless
Communications, December 2009}{}

\begin{abstract}
This paper presents the first in-depth analysis on the energy
efficiency of LT codes with Non Coherent M-ary Frequency Shift
Keying (NC-MFSK), known as \emph{green modulation}
\cite{JamshidTech_mod_2009}, in a proactive Wireless Sensor Network
(WSN) over Rayleigh flat-fading channels with path-loss. We describe
the proactive system model according to a pre-determined time-based
process utilized in practical sensor nodes. The present analysis is
based on realistic parameters including the effect of channel
bandwidth used in the IEEE 802.15.4 standard, and the active mode
duration. A comprehensive analysis, supported by some simulation
studies on the probability mass function of the LT code rate and
coding gain, shows that among uncoded NC-MFSK and various classical
channel coding schemes, the optimized LT coded NC-MFSK is the most
energy-efficient scheme for distance $d$ greater than the
pre-determined threshold level $d_T$, where the optimization is
performed over coding and modulation parameters. In addition,
although uncoded NC-MFSK outperforms coded schemes for $d < d_T$,
the energy gap between LT coded and uncoded NC-MFSK is negligible
for $d < d_T$ compared to the other coded schemes. These results
come from the flexibility of the LT code to adjust its rate to suit
instantaneous channel conditions, and suggest that LT codes are
beneficial in practical low-power WSNs with dynamic position sensor
nodes.
\end{abstract}

\begin{center}
  \centering{\bf{Index Terms}}

  \centering{\small Wireless sensor networks, energy efficiency, green modulation, LT codes.}
\end{center}

\section{Introduction}

Wireless Sensor Networks (WSNs) have been recognized as a new
generation of ubiquitous computing systems to support a broad range
of applications, including monitoring, health care and tracking
environmental pollution levels. Minimizing the total energy
consumption in both circuit components and RF signal transmission is
a crucial challenge in designing a WSN. Central to this study is to
find energy-efficient modulation and coding schemes in the physical
layer of a WSN to prolong the sensor lifetime
\cite{Cui_GoldsmithITWC0905, HowardEURASIP2006}. For this purpose,
energy-efficient modulation/coding schemes should be simple enough
to be implemented by state-of-the-art low-power technology, but
still robust enough to provide the desired service. Furthermore,
since sensor nodes frequently switch from sleep mode to active mode,
modulation and coding circuits should have fast start-up times
\cite{Wang_ISLPED2001} along with the capability of transmitting
packets during a pre-assigned time slot before new sensed packets
arrive. In addition, a WSN needs a powerful channel coding scheme
which protects transmitted data against the unpredictable and harsh
nature of channels. Finally, since coding increases the required
transmitted bandwidth, when considered independently of modulation,
the best tradeoff between energy-efficient modulation and coding for
a given transmission bandwidth should be considered as well. We
refer to these low-complexity and low-energy consumption approaches
in WSNs providing proper link reliability without increasing a given
transmission bandwidth as \emph{Green Modulation/Coding} (GMC)
schemes.

There have been several recent works on the energy efficiency of
various modulation and channel coding schemes in WSNs (see e.g.,
\cite{Cui_GoldsmithITWC0905, TangITWC0407, Chouhan_ITWC1009}). Tang
\emph{et al.} \cite{TangITWC0407} analyze the power efficiency of
Pulse Position Modulation (PPM) and Frequency Shift Keying (FSK) in
a WSN without considering the effect of channel coding. Under the
assumption of the non-linear battery model, reference
\cite{TangITWC0407} shows that FSK is more power-efficient than PPM
in sparse WSNs, while PPM may outperform FSK in dense WSNs.
Reference \cite{Sankarasubramaniam2003} investigates the energy
efficiency of BCH and convolutional codes with non-coherent FSK for
the optimal packet length in a point-to-point WSN. It is shown in
\cite{Sankarasubramaniam2003} that BCH codes can improve energy
efficiency compared to the convolutional code for optimal fixed
packet size. Reference \cite{Karvonen2004} analyzes the effect of
different linear block channel codes with FSK modulation on energy
consumptions of a low-power wireless embedded network when the
number of hops increases. Liang \emph{et al.} \cite{LiangPACRIM2007}
investigate the energy efficiency of uncoded NC-MFSK modulation
scheme in a multiple-access WSN over Rayleigh fading channels, where
multiple senders transmit their data to a central node in a
Frequency-Division Multiple Access (FDMA) fashion. Reference
\cite{Hanzo_2009} presents the hardware implementation of the
Forward Error Correction (FEC) encoder in IEEE 802.15.4 WSNs, which
employs parallel just-in-time processing to achieve a low processing
latency and energy consumption.

Most of the pioneering works on energy-efficient modulation/coding,
including research in \cite{TangITWC0407, Qu_ITSP0908, Shen2008,
Sankarasubramaniam2003, Karvonen2004}, has focused only on
minimizing the energy consumption of transmitting one bit, ignoring
the effect of bandwidth and transmission time duration. In a
practical WSN however, it is shown that minimizing the total energy
consumption depends strongly on the active mode duration and the
channel bandwidth. References \cite{JamshidTech_mod_2009,
Cui_GoldsmithITWC0905} and \cite{Chouhan_ITWC1009} address this
issue in a point-to-point WSN, where a sensor node transmits an
equal amount of data per time unit to a designated sink node. In
\cite{Cui_GoldsmithITWC0905}, the authors consider the optimal
energy consumption per information bit as a function of modulation
and coding parameters in a WSN over Additive White Gaussian Noise
(AWGN) channels with path-loss. It is shown in
\cite{Cui_GoldsmithITWC0905} that uncoded MQAM is more
energy-efficient than uncoded MFSK for short-range applications. For
higher distance, however, using convolutional coded MFSK over AWGN
is desirable. This line of work is further extended in
\cite{Chouhan_ITWC1009} by evaluating the energy consumption per
information bit of a WSN for Reed Solomon (RS) Codes and various
modulation schemes over AWGN channels with path-loss. Also, the
impact of different transmission distances on the energy consumption
per information bit is investigated in \cite{Chouhan_ITWC1009}. In
\cite{Cui_GoldsmithITWC0905} and \cite{Chouhan_ITWC1009}, the
authors do not consider the effect of multi-path fading. Reference
\cite{JamshidTech_mod_2009} addresses this problem in a similar WSN
model as \cite{Cui_GoldsmithITWC0905} and \cite{Chouhan_ITWC1009},
and shows that among various sinusoidal carrier-based modulation
schemes, Non-Coherent M-ary Frequency Shift Keying (NC-MFSK) with
small order of constellation size $M$ can be considered the most
energy-efficient modulation in proactive WSNs over Rayleigh and
Rician fading channels. However, no channel coding scheme was
considered in \cite{JamshidTech_mod_2009}.

More recently, the attention of researchers has been drawn to
deploying rateless codes (e.g., Luby Transform (LT) code
\cite{LubyFOCS2002}) in WSNs due to the outstanding advantages of
these codes in \emph{erasure channels}. For instance in
\cite{Eckford_ICC2006}, the authors present a scheme for cooperative
error control coding using rateless and Low-Density Generator-Matrix
(LDGM) codes in a multiple relay WSN. However, investigating the
energy efficiency of rateless codes in WSNs with green modulations
over realistic fading channel models has received little attention.
To the best of our knowledge, there is no existing analysis on the
energy efficiency of rateless coded modulation that considers the
effect of channel bandwidth and active mode duration on the total
energy consumption in a practical proactive WSN. This paper
addresses this problem and presents the first in-depth analysis of
the energy efficiency of LT codes with NC-MFSK (known as green
modulation) as described in \cite{JamshidTech_mod_2009}. The present
analysis is based on a realistic model in proactive WSNs operating
in a Rayleigh flat-fading channel with path-loss. In addition, we
obtain numerically the probability mass function of the LT code rate
and the corresponding coding gain, and study their effects on the
energy efficiency of the WSN. This study uses the classical BCH and
convolutional codes (as reference codes), utilized in IEEE
standards, for comparative evaluation. Experimental results show
that the optimized LT coded NC-MFSK is the most energy-efficient
scheme for distance $d$ greater than the threshold level $d_T$. In
addition, although uncoded NC-MFSK outperforms coded schemes for $d
< d_T$, the energy gap between LT coded and uncoded NC-MFSK is
negligible for $d < d_T$ compared to the other coded schemes. This
result comes from the simplicity and flexibility of the LT codes,
and suggests that LT codes are beneficial in practical low-power
WSNs with dynamic position sensor nodes.

The rest of the paper is organized as follows. In Section
\ref{System_model}, the proactive system model over a realistic
wireless channel model is described. The energy consumption of
uncoded NC-MFSK modulation scheme is analyzed in Section
\ref{uncoded_MFSK}. Design of LT codes and the energy efficiency of
the LT coded NC-MFSK are presented in Section \ref{analysis_Ch4}. In
addition, the energy efficiency of some classical channel codes are
studied in this section. Section \ref{simulation_Ch5} provides some
numerical evaluations using realistic models to confirm our
analysis. Also, some design guidelines for using LT codes in
practical WSN applications are presented. Finally in Section
\ref{conclusion_Ch6}, an overview of the results and conclusions are
presented.

For convenience, we provide a list of key mathematical symbols used
in this paper in Table I.

\begin{table}
   \label{table234}
\caption{List of Notations } \centering
  \begin{tabular}{|l|l|}
  \hline
  $B$  &  Channel bandwidth\\
  $d$  & Transmission distance\\
  $\mathcal{E}_t$  & Energy of uncoded transmitted signal\\
  $\mathcal{E}_{N}$  & Total energy consumption for uncoded case\\
  $h_{i}$  &  Fading channel coefficient for symbol $i$\\
  $\mathcal{L}_d$      &  Channel gain factor with distance $d$\\
  $M$  & Constellation size\\
  $N$    & Number of sensed message\\
  $n$  & Codeword block length\\
  $\mathcal{O}(x)$ & Output-node degree distribution\\
  $\mathcal{P}_c$  & Circuit power consumption\\
  $\mathcal{P}_t$  & Power of transmitted signal\\
  $P_b$   & Bit error rate\\
  $P_R(\ell)$  & pmf of LT code rate\\
  $R_c$  &  Code rate\\
  $T_{ac}$ & Active mode duration\\
  $T_s$   &  Symbol duration\\
  $\eta$  & Path-loss exponent\\
  $\Omega$ & $\mathbb{E}\left[\vert h_{i} \vert^2\right]$ \\
  $\gamma_{i}$  &  Instantaneous SNR \\
  $\Upsilon_{c}$ & Coding gain\\
  \hline
  \end{tabular}
\end{table}

\section{System Model and Assumptions}\label{System_model}

\begin{figure}[t]
\centerline{\psfig{figure=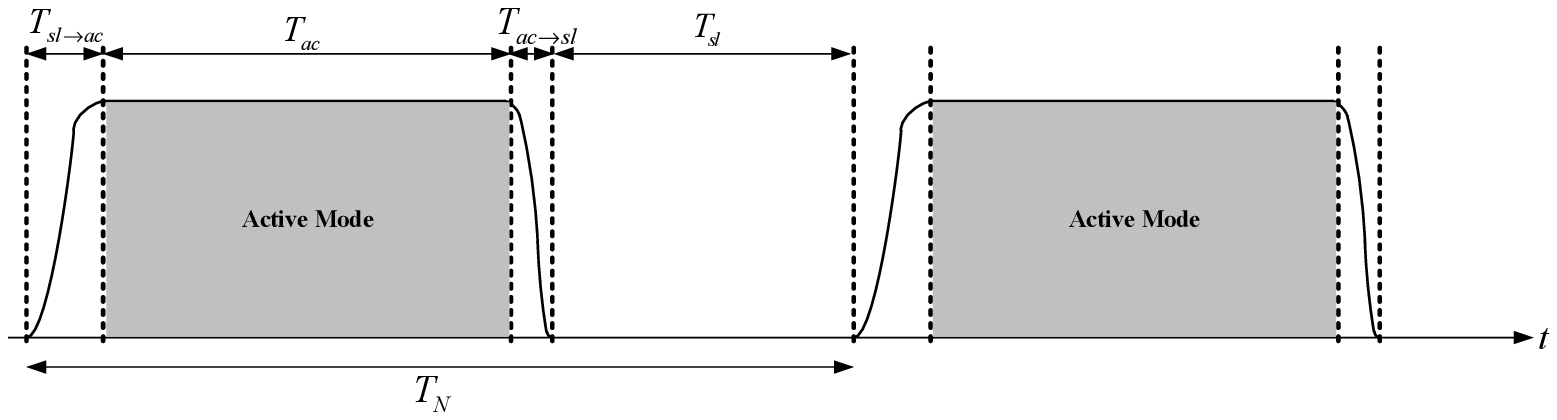,width=4.95in}} \caption{A
practical multi-mode operation in a proactive WSN. } \label{fig:
Time-Basis}
\end{figure}

In this work, we consider a \emph{proactive} wireless sensor system,
in which a sensor node continuously samples the environment and
transmits an equal amount of data per time unit to a designated sink
node. Such a proactive sensor system is typical of many
environmental applications such as sensing temperature, humidity,
level of contamination, etc \cite{Cordeiro_Book2006}. For this
proactive system, the sensor and sink nodes synchronize with one
another and operate in a real time-based process as depicted in Fig
\ref{fig: Time-Basis}. During \emph{active mode} duration $T_{ac}$,
the analog signal sensed by the sensor is first digitized by an
Analog-to-Digital Converter (ADC), and an $N$-bit message sequence
$\mathcal{M}_N\triangleq (m_1,m_2,...,m_N)$ is generated, where $N$
is assumed to be fixed, and $m_i \in \{0,1 \}$, $i=1,2,...,N$. The
bit stream is then sent to the channel encoder. The encoding process
begins by dividing the uncoded message $\mathcal{M}_N$ into blocks
of equal length denoted by $\mathcal{B}_j \triangleq
(m_{_{(j-1)k+1}},...,m_{jk})$, $j=1,...,\frac{N}{k}$, where $k$ is
the length of any particular $\mathcal{B}_j$, and $N$ is assumed to
be divisible by $k$. Each block $\mathcal{B}_j$ is encoded by a
pre-determined channel coding scheme to generate a coded bit stream
$\mathcal{C}_j \triangleq (a_{_{(j-1)n+1}},...,a_{jn})$,
$j=1,...,\frac{N}{k}$, with block length $n$, where $n$ is either a
fixed value (e.g., for block and convolutional codes) or a random
variable (e.g., for LT codes).

The coded stream is then modulated by an NC-MFSK scheme and
transmitted to a designated sink node. Finally, the sensor node
returns to \emph{sleep mode}, and all the circuits of the
transceiver are shutdown for sleep mode duration $T_{sl}$ for energy
saving. We denote $T_{tr}$ as the \emph{transient mode} duration
consisting of the switching time from sleep mode to active mode
(i.e., $T_{sl \rightarrow ac}$) plus the switching time from active
mode to sleep mode (i.e., $T_{ac \rightarrow sl}$), where $T_{ac
\rightarrow sl}$ is short enough compared to $T_{sl \rightarrow ac}$
to be negligible. Furthermore, when the sensor switches from sleep
mode to active mode to send data, a significant amount of power is
consumed for starting up the transmitter, while the power
consumption during $T_{ac \rightarrow sl}$ is negligible. Under the
above considerations, the sensor/sink nodes have to process one
entire $N$-bit message $\mathcal{M}_N$ during $0 \leq T_{ac} \leq
T_N-T_{tr}$, before a new sensed packet arrives, where $T_N
\triangleq T_{tr}+T_{ac}+T_{sl}$ is fixed, and $T_{tr} \approx T_{sl
\rightarrow ac}$.

Since sensor nodes in a typical WSN are densely deployed, the
distance between nodes is normally short. Thus, the circuit power
consumption in a WSN is comparable to the output transmit power
consumption. We denote the total circuit power consumption as
$\mathcal{P}_c \triangleq \mathcal{P}_{ct}+\mathcal{P}_{cr}$, where
$\mathcal{P}_{ct}$ and $\mathcal{P}_{cr}$ represent the circuit
power consumptions for sensor and sink nodes, respectively. In
addition, the power consumption of RF signal transmission in the
sensor node is denoted by $\mathcal{P}_t$. Taking these into
account, the total energy consumption during the active mode period,
denoted by $\mathcal{E}_{ac}$, is given by
$\mathcal{E}_{ac}=(\mathcal{P}_c+\mathcal{P}_t)T_{ac}$. Also, the
energy consumption in the sleep mode period, denoted by
$\mathcal{E}_{sl}$, is given by
$\mathcal{E}_{sl}=\mathcal{P}_{sl}T_{sl}$, where $\mathcal{P}_{sl}$
is the corresponding power consumption. It is worth mentioning that
during the sleep mode interval, the \emph{leakage current} coming
from the CMOS circuits embedded in the sensor node is a dominant
factor in $\mathcal{P}_{sl}$. Clearly, higher sleep mode duration
increases the energy consumption $\mathcal{E}_{sl}$ due to
increasing leakage current as well as $T_{sl}$. Present state-of-the
art technology aims to keep a low sleep mode leakage current no
larger than the battery leakage current, which results in
$\mathcal{P}_{sl}$ much smaller than the power consumption in active
mode \cite{Mingoo2007}. For this reason, we assume that
$\mathcal{P}_{sl}\approx 0$. As a result, we have the following
definition.

\begin{defi}\label{Definition01}
\textbf{(Performance Metric):} The energy efficiency, referred to as
the performance metric of the proposed WSN, can be measured by the
total energy consumption in each period $T_N$ corresponding to
$N$-bit message $\mathcal{M}_N$ as follows:
\begin{equation}\label{total_energy1}
\mathcal{E}_N =
(\mathcal{P}_c+\mathcal{P}_t)T_{ac}+\mathcal{P}_{tr}T_{tr},
\end{equation}
where $\mathcal{P}_{tr}$ is the circuit power consumption during the
transient mode period.
\end{defi}
We use (\ref{total_energy1}) to investigate and compare the energy
efficiency of uncoded and coded NC-MFSK for various channel coding
schemes.

\textbf{Channel Model:} The choice of low transmission power in WSNs
results in several consequences to the channel model. It is shown by
Friis \cite{Friis1946} that a low transmission power implies a small
range. For short-range transmission scenarios, the root mean square
(rms) delay spread is in the range of nanoseconds
\cite{Karl_Book2005} which is small compared to symbol durations for
modulated signals. For instance, the channel bandwidth and the
corresponding symbol duration considered in the IEEE 802.15.4
standard are $B=62.5$ KHz and $T_s=16~\mu$s, respectively \cite[p.
49]{IEEE_802_15_4_2006}, while the rms delay spread in indoor
environments are in the range of 70-150 ns \cite{Barclay2003}. Thus,
it is reasonable to expect a flat-fading channel model for WSNs. In
addition, many transmission environments include significant
obstacle and structural interference by obstacles (such as wall,
doors, furniture, etc), which leads to reduced Line-Of-Sight (LOS)
components. This behavior suggests a Rayleigh fading channel model.
Under the above considerations, the channel model between the sensor
and sink nodes is assumed to be Rayleigh flat-fading with path-loss,
which is a feasible model in static WSNs \cite{TangITWC0407,
Qu_ITSP0908}. For this model, we assume that the channel is constant
during the transmission of a codeword, but may vary from one
codeword to another. We denote the fading channel coefficient
corresponding to an arbitrary transmitted symbol $i$ as $h_{i}$,
where the amplitude $\big\vert h_{i} \big\vert$ is Rayleigh
distributed with probability density function (pdf) given according
to $f_{\vert h_{i}
\vert}(r)=\frac{2r}{\Omega}e^{-\frac{r^2}{\Omega}},~r \geq 0$, where
$\Omega \triangleq \mathbb{E}\left[\vert h_{i} \vert^2\right]$
(e.g., pp. 767-768 of \cite{Proakis2001}). This results in $\vert
h_{i} \vert^2$ being \emph{chi-square} distributed with 2 degrees of
freedom, where $f_{\vert h_{i}
\vert^2}(r)=\frac{1}{\Omega}e^{\frac{-r}{\Omega}},~r \geq 0$.

To model the path-loss of a link where the transmitter and receiver
are separated by distance $d$, let denote $\mathcal{P}_t$ and
$\mathcal{P}_r$ as the transmitted and the received signal powers,
respectively. For a $\eta^{th}$-power path-loss channel, the channel
gain factor is given by $\mathcal{L}_d \triangleq
\frac{\mathcal{P}_t}{\mathcal{P}_r}=M_ld^\eta \mathcal{L}_1$, where
$M_l$ is the gain margin which accounts for the effects of hardware
process variations, background noise and $\mathcal{L}_1 \triangleq
\frac{(4 \pi)^2}{\mathcal{G}_t \mathcal{G}_r \lambda^2}$ is the gain
factor at $d=1$ meter which is specified by the transmitter and
receiver antenna gains $\mathcal{G}_t$ and $\mathcal{G}_r$, and
wavelength $\lambda$ (e.g., \cite{Cui_GoldsmithITWC0905},
\cite{Qu_ITSP0908}, Ch. 4 of \cite{Rappaport2002} and
\cite{MinISPED2002}). As a result, when both fading and path-loss
are considered, the instantaneous channel coefficient corresponding
to an arbitrary symbol $i$ becomes $G_{i} \triangleq
\frac{h_{i}}{\sqrt{\mathcal{L}_d}}$. Denoting $x_i(t)$ as the RF
transmitted signal with energy $\mathcal{E}_{t}$, the received
signal at the sink node is given by
$y_{i}(t)=G_{i}x_{i}(t)+n_{i}(t)$, where $n_{i}(t)$ is AWGN at the
sink node with two-sided power spectral density given by
$\frac{N_{0}}{2}$. Under the above considerations, the instantaneous
Signal-to-Noise Ratio (SNR), denoted by $\gamma_{i}$, corresponding
to an arbitrary symbol $i$ can be computed as
$\gamma_{i}=\frac{\vert G_{i}\vert^2 \mathcal{E}_t}{N_0}$. Under the
assumption of a Rayleigh fading channel model, $\gamma_{i}$ is
chi-square distributed with 2 degrees of freedom and with pdf
$f_{\gamma}(\gamma_{i})=\frac{1}{\bar{\gamma}}e^{
-\frac{\gamma_{i}}{\bar{\gamma}}}$, where $\bar{\gamma} \triangleq
\mathbb{E}[\vert
G_{i}\vert^2]\frac{\mathcal{E}_t}{N_0}=\frac{\Omega}{\mathcal{L}_d}\frac{\mathcal{E}_t}{N_0}$
denotes the average received SNR.

\section{Energy Consumption of Uncoded NC-MFSK Modulation}\label{uncoded_MFSK}
We first consider uncoded MFSK modulation in the proposed proactive
WSN, where $M$ orthogonal carriers can be mapped into $b \triangleq
\log_{2}M$ bits. Denoting $\mathcal{E}_t$ as the uncoded MFSK
transmit energy per symbol with symbol duration $T_s$, the
transmitted signal from the sensor node is given by
$x_{i}(t)=\sqrt{\frac{2\mathcal{E}_t}{T_s}}\cos(2\pi(f_0+i\Delta
f)t),~i=0,1,...M-1$, where $f_0$ is the first carrier frequency in
the MFSK modulator and $\Delta f=\frac{1}{T_s}$ is the minimum
carrier separation in the non-coherent case. Thus, the channel
bandwidth is obtained as $B \approx M\times\Delta f $, which is
assumed to be a fixed value. Since we have $b$ bits during each
symbol period $T_{s}$, we can write
\begin{equation}\label{active1}
T_{ac}=\dfrac{N}{b}T_{s}=\dfrac{MN}{B\log_2 M}.
\end{equation}
Recalling that $B$ and $N$ are fixed, an increase $M$ results in an
increase in $T_{ac}$. However, as illustrated in Fig. \ref{fig:
Time-Basis}, the maximum value for $T_{ac}$ is bounded by
$T_N-T_{tr}$. Thus, the maximum constellation size $M$, denoted by
$M_{max}\triangleq 2^{b_{max}}$, for uncoded MFSK is calculated by
$\frac{2^{b_{max}}}{b_{max}}=\frac{B}{N}(T_{N}-T_{tr})$. It is shown
in \cite{JamshidTech_mod_2009} that the transmit energy consumption
per each symbol for an uncoded NC-MFSK is obtained as
\begin{eqnarray}
\mathcal{E}_t &\triangleq& \mathcal{P}_t T_s \approx \left[\left(
1-(1-P_s)^{\frac{1}{M-1}}\right)^{-1}-2 \right]\dfrac{\mathcal{L}_d
N_0}{\Omega}\\
\label{deriv1}&\stackrel{(a)}{=}& \left[\left(
1-\left(1-\frac{2(M-1)}{M}P_b\right)^{\frac{1}{M-1}}\right)^{-1}-2
\right]\dfrac{\mathcal{L}_d N_0}{\Omega},
\end{eqnarray}
where $(a)$ comes from the fact that the relationship between the
average Symbol Error Rate (SER) $P_s$ and the average Bit Error Rate
(BER) $P_b$ of MFSK is given by $P_s=\frac{2(M-1)}{M}P_b$ \cite[p.
262]{Proakis2001}. Using (\ref{active1}), the output energy
consumption of transmitting $N$-bit during $T_{ac}$ of an uncoded
NC-MFSK is then computed as
\begin{equation}\label{energy_trans1}
\mathcal{P}_t T_{ac}= \dfrac{T_{ac}}{T_s}\mathcal{E}_t \approx
\left[\left(
1-\left(1-\frac{2(M-1)}{M}P_b\right)^{\frac{1}{M-1}}\right)^{-1}-2
\right]\dfrac{\mathcal{L}_d N_0}{\Omega} \dfrac{N}{\log_2 M}.
\end{equation}
For the sensor node with uncoded MFSK, we denote the power
consumption of frequency synthesizer, filters and power amplifier as
$\mathcal{P}_{Sy}$, $\mathcal{P}_{Filt}$ and $\mathcal{P}_{Amp}$,
respectively. In this case, the power consumption of the sensor
circuitry with uncoded MFSK can be obtained as
\begin{equation}\label{sensor_power}
\mathcal{P}_{ct}=\mathcal{P}_{Sy}+\mathcal{P}_{Filt}+\mathcal{P}_{Amp},
\end{equation}
where $\mathcal{P}_{Amp}=\alpha \mathcal{P}_{t}$, in which $\alpha$
is determined based on the type (or equivalently drain efficiency)
of power amplifier. For instance, for a class B power amplifier,
$\alpha=0.33$ \cite{Cui_GoldsmithITWC0905}, \cite{TangITWC0407}. It
is shown in \cite{JamshidTech_mod_2009} that the power consumption
of the sink circuitry with uncoded NC-MFSK scheme can be obtained as
\begin{equation}\label{sink_power}
\mathcal{P}_{cr}=\mathcal{P}_{LNA}+M\times(\mathcal{P}_{Filr}+\mathcal{P}_{ED})+\mathcal{P}_{IFA}+\mathcal{P}_{ADC},
\end{equation}
where $\mathcal{P}_{LNA}$, $\mathcal{P}_{Filr}$, $\mathcal{P}_{ED}$,
$\mathcal{P}_{IFA}$ and $\mathcal{P}_{ADC}$ denote the power
consumption of Low-Noise Amplifier (LNA), filters, envelop detector,
IF amplifier and ADC, respectively. In addition, it is shown that
the power consumption during transition mode period $T_{tr}$ is
governed by the frequency synthesizer \cite{Wang_ISLPED2001}. Thus,
the energy consumption during $T_{tr}$ is obtained as
$\mathcal{P}_{tr}T_{tr}=1.75 \mathcal{P}_{Sy}T_{tr}$
\cite{Karl_Book2005}. Substituting
(\ref{active1})-(\ref{sink_power}) in (\ref{total_energy1}), the
total energy consumption of an uncoded NC-MFSK scheme for
transmitting $N$-bit information in each period $T_N$, under the
constraint $M \leq M_{max}$ and for a given $P_b$ is obtained as
\begin{equation}\label{energy_totFSK}
\mathcal{E}_N = (1+\alpha) \left[\left(
1-\left(1-\dfrac{2(M-1)}{M}P_b\right)^{\frac{1}{M-1}}\right)^{-1}-2
\right]\dfrac{\mathcal{L}_d N_0}{\Omega} \dfrac{N}{\log_2 M}+
(\mathcal{P}_{c}-\mathcal{P}_{Amp})\dfrac{MN}{B\log_{2}M}+1.75
\mathcal{P}_{Sy}T_{tr}.
\end{equation}

It is shown in \cite{JamshidTech_mod_2009} that the above uncoded
NC-MFSK is more energy-efficient than other sinusoidal carrier-based
modulation schemes, and is a good option for low-power and low data
rate WSN applications. For energy optimal designs, however, the
impact of channel coding on the energy efficiency of the proposed
WSN must be considered as well. It is a well known fact that channel
coding is a classical approach used to improve the link reliability
along with the transmitter energy saving due to providing the coding
gain \cite{Proakis2001}. However, the energy saving comes at the
cost of extra energy spent in transmitting the redundant bits in
codewords as well as the additional energy consumption in the
process of encoding/decoding. For a specific transmission distance
$d$, if these extra energy consumptions outweigh the transmit energy
saving due to the channel coding, the coded system would not be
energy-efficient compared with an uncoded system. In the subsequent
sections, we will argue the above problem and determine at what
distance use of specific channel coding becomes energy-efficient
compared to uncoded systems. In particular, we will show in Section
\ref{simulation_Ch5} that the LT coded NC-MFSK surpasses this
distance constraint in the proposed WSN.

\section{Energy Consumption Analysis of LT Coded NC-MFSK}\label{analysis_Ch4}
In this section, we present the first in-depth analysis on the
energy efficiency of LT coded NC-MFSK for the proposed proactive
WSN. To get more insight into how channel coding affects the circuit
and RF signal energy consumptions in the system, we modify the
energy concepts in Section \ref{uncoded_MFSK}, in particular, the
total energy consumption expression in (\ref{energy_totFSK}) based
on the coding gain and code rate. To address this problem and for
the purpose of comparative evaluation, we first start with classical
BCH and convolutional channel codes (referred to as fixed-rate
codes), that are widely utilized in IEEE standards \cite{IEEE802.15,
IEEE802.16}. We further present the first study on the tradeoff
between LT code rate and coding gain required to achieve a certain
BER, and the effect of this tradeoff on the total energy consumption
of LT coded NC-MFSK for different transmission distances\footnote{In
the sequel and for simplicity of notation, we use the superscripts
`BC', `CC' and `LT' for BCH, convolutional and LT codes,
respectively.}.

\textbf{BCH Codes:} In the BCH$(n,k,t)$ code with up to $t$-error
correction capability, each $k$-bit message $\mathcal{B}_j \in
\mathcal{M}_N$ is encoded into $2^{nR^{BC}_c}$ valid codewords
$\mathcal{C}_j \triangleq (a_{_{(j-1)n+1}},...,a_{jn})$,
$j=1,...,\frac{N}{k}$ with block length $n$, where $R^{BC}_c
\triangleq \frac{k}{n}$ is the BCH code rate. For this code,
$n=2^m-1$ bits, where $m \geq 3$, $t < 2^{m-1}$ and the number of
parity-check bits is upper bounded by $n-k \leq mt$. It is worth
mentioning that the number of transmitted bits in each period $T_N$
is increased from $N$-bit uncoded message to
$\frac{N}{R^{BC}_c}=\frac{N}{k}n$ bits coded one. To compute the
total energy consumption of coded scheme, we use the fact that
channel coding can reduce the required average SNR value to achieve
a given BER. Taking this into account, the proposed WSN with BCH
codes benefits in transmission energy saving specified by
$\mathcal{E}_{t}^{BC}=\frac{\mathcal{E}_{t}}{\Upsilon_{c}^{BC}}$,
where $\Upsilon_{c}^{BC} \geq 1$ is the coding
gain\footnote{Denoting $\bar{\gamma}
=\frac{\Omega}{\mathcal{L}_d}\frac{\mathcal{E}_t}{N_0}$ and
$\bar{\gamma}^{BC}
=\frac{\Omega}{\mathcal{L}_d}\frac{\mathcal{E}^{BC}_t}{N_0}$ as the
average SNR of uncoded and BCH coded schemes, respectively, the BCH
\emph{coding gain} (expressed in dB) is defined as the difference
between the values of $\bar{\gamma}$ and $\bar{\gamma}^{BC}$
required to achieve a certain BER, where
$\mathcal{E}_{t}^{BC}=\frac{\mathcal{E}_{t}}{\Upsilon_{c}^{BC}}$.}
of BCH coded NC-MFSK. Table II displays the coding gain of some
BCH$(n,k,t)$ codes with NC-MFSK scheme over a Rayleigh flat-fading
channel for different values of $M$ and given
$P_b=(10^{-3},10^{-4})$. For these results, a hard-decision decoding
algorithm is considered. It is seen from Table II that there is a
tradeoff between coding gain and decoder complexity. In fact,
achieving a higher coding gain for given $M$, requires a more
complex decoding process, (i.e., higher $t$) with more circuit power
consumption.

\begin{figure}[t]
\centerline{\psfig{figure=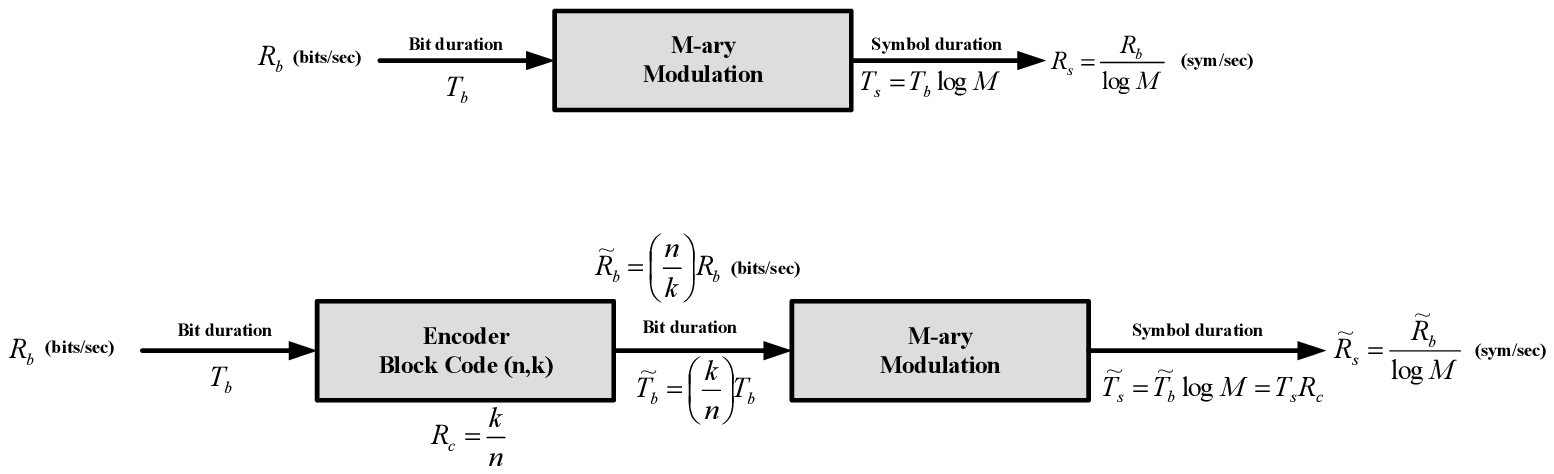,width=5.95in}}
\caption{A schematic diagram of the bandwidth expansion in an
arbitrary block code$(n,k)$ with M-ary modulation scheme. }
\label{fig: Bandwidth_Expansion}
\end{figure}

It should be noted that the cost of the energy savings of using BCH
codes is the bandwidth expansion $\frac{B}{R^{BC}_{c}}$ as depicted
in Fig. \ref{fig: Bandwidth_Expansion}. In order to keep the
bandwidth of the coded system the same as that of the uncoded case,
we must keep the information transmission rate constant, i.e., the
symbol duration $T_{s}$ of uncoded and coded NC-MFSK would be the
same. Thus, we can drop the superscript ``BC'' in $T_s$ for the
coded case. However, the active mode duration increases from
$T_{ac}=\frac{N}{b}T_{s}$ in the uncoded system to
\begin{equation}\label{BCH_active}
T_{ac}^{BC}=\frac{N}{bR_{c}^{BC}}T_{s}=\frac{T_{ac}}{R_{c}^{BC}}
\end{equation}
for the BCH coded case. Thus, one would assume that the total time
$T_N$ increases to $\frac{T_N}{R^{BC}_c}$ for the coded scenario. It
should be noted that the maximum constellation size $M$, denoted by
$M_{max}\triangleq 2^{b_{max}}$, for the coded NC-MFSK is calculated
by $\frac{2^{b_{max}}}{b_{max}}=\frac{B
R_{c}^{BC}}{N}(\frac{T_{N}}{R_{c}^{BC}}-T_{tr})$, which is
approximately the same as that of the uncoded case.

\begin{table}
\label{table01} \caption{Coding gain ($dB$) of BCH$(n,k,t)$ and
convolutional coded NC-MFSK over Rayleigh flat-fading channel for
BER$=(10^{-3},10^{-4})$.} \centering
  \begin{tabular}{|l|c| c c c c c c|}
  \hline
  BCH Code  & $R^{BC}_c$ & M=2  & M=4  & M=8  &  M=16  &  M=32  &  M=64
  \\
  \hline
   BCH $(7,4,1)$   & 0.571 &   $(2.5,2.8)$   &   $(0.3,0.4)$    &    $(0.1,0.2)$  &  $(0.0,0.0)$   &  $(0.0,0.0)$   &  $(0.0,0.0)$   \\
  \hline
   BCH $(15,11,1)$ & 0.733 &   $(1.4,1.6)$   &   $(0.2,0.3)$    &    $(0.0,0.0)$  &  $(0.0,0.0)$   &  $(0.0,0.0)$   &  $(0.0,0.0)$   \\
   \hline
   BCH $(15,7,2)$  & 0.467 &  $(2.4,3.3)$   &   $(2.0,2.3)$    &    $(0.8,1.0)$  &  $(0.3,0.4)$   &  $(0.0,0.0)$   &  $(0.0,0.0)$   \\
   \hline
   BCH $(15,5,3)$  & 0.333 &  $(4.1,4.6)$   &   $(2.7,2.9)$    &    $(2.0,2.1)$  &  $(1.5,1.60)$  &  $(0.7,0.8)$   &  $(0.2,0.2)$   \\
   \hline
   BCH $(31,26,1)$ & 0.839 &  $(1.2,1.5)$   &   $(0.2,0.2)$    &    $(0.0,0.0)$  &  $(0.0,0.0)$   &  $(0.0,0.0)$   &  $(0.0,0.0)$   \\
   \hline
   BCH $(31,21,2)$ & 0.677 &  $(2.3,2.9)$   &   $(1.7,2.0)$    &    $(0.7,0.8)$  &  $(0.2,0.2)$   &  $(0.0,0.0)$   &  $(0.0,0.0)$   \\
   \hline
   BCH $(31,16,3)$ & 0.516 &  $(2.9,3.1)$   &   $(2.1,2.2)$    &    $(1.5,1.6)$  &  $(1.3,1.4)$   &  $(0.6,0.7)$   &  $(0.1,0.1)$   \\
   \hline
   BCH $(31,11,5)$ & 0.355 &  $(4.1,4.4)$   &   $(3.5,4.2)$    &    $(2.2,2.3)$  &  $(2.0,2.1)$   &  $(1.8,2.0)$   &  $(1.1,1.3)$   \\
   \hline
   BCH $(31,6,7)$  & 0.194 &  $(5.4,5.9)$   &   $(4.3,4.8)$    &    $(3.5,3.8)$  &  $(3.2,3.3)$   &  $(2.7,2.8)$   &  $(2.3,2.4)$   \\
   \hline
   \hline
   Convolutional Code  & $R^{CC}_c$ & M=2  & M=4  & M=8  &  M=16  &  M=32  &  M=64\\
  \hline
  trel$(6,[53~~75])$    & 0.500 &   $(3.8,4.6)$  &   $(2.7,3.1)$  &    $(2.1,2.3)$  &  $(1.8,2.0)$   &  $(1.4,1.5)$   &  $(1.4,1.4)$   \\
  \hline
   trel$(7,[133~~171])$  & 0.500 &   $(4.0,4.7)$  &   $(3.0,3.5)$  &    $(2.2,2.4)$  &  $(1.8,2.0)$   &  $(1.5,1.6)$   &  $(1.4,1.5)$   \\
  \hline
  trel$(7,[133~~165~~171])$  & 0.333 &   $(5.7,6.4)$  &   $(4.8,5.1)$  &    $(3.7,3.9)$  &  $(3.1,3.3)$   &  $(2.7,2.8)$   &  $(2.5,2.6)$   \\
   \hline
   trel$([4~3],[4~5~17;7~4~2])$  & 0.667 &   $(2.2,2.6)$  &   $(1.5,1.7)$  &    $(0.9,1.1)$  &  $(0.6,0.6)$   &  $(0.5,0.5)$   &  $(0.5,0.5)$   \\
   \hline
   trel$([5~4],[23~35~0;0~5~13])$  & 0.667 &   $(2.9,3.5)$  &   $(1.9,2.4)$  &    $(1.4,1.8)$  &  $(1.1,1.2)$   &  $(0.8,0.9)$   &  $(0.7,0.8)$   \\
   \hline
   \end{tabular}
\end{table}

Now, we are ready to compute the total energy consumption in the
case of BCH coded NC-MFSK. Since BCH codes are implemented using
Linear-Feedback Shift Register (LFSR) circuits, the BCH encoder can
be assumed to have negligible energy consumption. Thus, the energy
cost of the sensor circuity with BCH coded NC-MFSK scheme is
approximately the same as that of uncoded one. Also, the energy
consumption of an BCH decoder is negligible compared to the other
circuit components in the sink node, as shown in Appendix
\ref{decoder}. Substituting (\ref{deriv1}) in
$\mathcal{E}_{t}^{BC}=\frac{\mathcal{E}_{t}}{\Upsilon_{c}^{BC}}$,
and using (\ref{sensor_power}), (\ref{sink_power}) and
(\ref{BCH_active}), the total energy consumption of transmitting
$\frac{N}{R_c^{BC}}$ bits in each period $\frac{T_N}{R_C^{BC}}$ for
a BCH coded NC-MFSK scheme, and for a given $P_b$ is obtained as
\begin{eqnarray}
\notag \mathcal{E}^{BC}_N &=& (1+\alpha) \left[\left(
1-\left(1-\dfrac{2(M-1)}{M}P_b\right)^{\frac{1}{M-1}}\right)^{-1}-2
\right]\dfrac{\mathcal{L}_d N_0}{\Omega \Upsilon_{c}^{BC}} \dfrac{N}{R_{c}^{BC}\log_2 M}+\\
\label{energy_totalcodedFSK}&&(\mathcal{P}_{c}-\mathcal{P}_{Amp})\dfrac{MN}{BR_{c}^{BC}
\log_{2}M}+ 1.75 \mathcal{P}_{Sy}T_{tr},
\end{eqnarray}
under the constraint $M \leq M_{max}$.

\textbf{Convolutional Codes:} A convolutional code $(n,k,L)$ is
commonly specified by the number of input bits $k$, the number of
output bits $n$, and the constraint length\footnote{The constraint
length $L$ represents the number of bits in the encoder memory that
affect the generation of the $n$ output bits.} $L$. As with BCH
codes, the rate of a convolutional code is given by the ratio
$R^{CC}_c \triangleq \frac{k}{n}$. Since $k$ and $n$ are small
integers (typically from 1 to 8) with $k < n$, the convolutional
encoder is extremely simple to implement and can be assumed to have
negligible energy consumption. With a similar argument as for BCH
codes, the convolutional coded NC-MFSK provides an energy saving
compared to the uncoded system, which is specified by
$\mathcal{E}_{t}^{CC}=\frac{\mathcal{E}_t}{\Upsilon_{c}^{CC}}$.
Table II gives the coding gain $\Upsilon_{c}^{CC}$ of some practical
convolutional codes used in IEEE standards with an NC-MFSK
modulation scheme, over a Rayleigh flat-fading channel for different
constellation size $M$ and given $P_b=(10^{-3},10^{-4})$. For these
results, a hard-decision Viterbi decoding algorithm is considered.
It is seen from Table II that for a given $M$, the convolutional
codes with lower rates and higher constraint lengths achieve greater
coding gains. Also, in contrast to \cite{Cui_GoldsmithITWC0905},
where the authors assume a fixed convolutional coding gain for every
value of $M$, it is observed that the coding gain of convolutional
coded NC-MFSK is a monotonically decreasing function of $M$.

It should be noted that the energy efficiency analysis presented for
BCH codes, in particular deriving the total energy consumption
$\mathcal{E}_{N}^{CC}$, is valid for convolution codes. Thus, by
substituting  (\ref{deriv1}) in
$\mathcal{E}_{t}^{CC}=\frac{\mathcal{E}_{t}}{\Upsilon_{c}^{CC}}$,
and using (\ref{sensor_power}), (\ref{sink_power}) and
$T_{ac}^{CC}=\frac{N}{bR_{c}^{CC}}T_{s}$, the total energy
consumption of transmitting $\frac{N}{R_c^{CC}}$ bits in each period
$\frac{T_N}{R^{CC}_c}$ for a convolutional coded NC-MFSK scheme,
achieving a certain $P_b$, is obtained as
\begin{eqnarray}
\notag \mathcal{E}^{CC}_N &=& (1+\alpha) \left[\left(
1-\left(1-\dfrac{2(M-1)}{M}P_b\right)^{\frac{1}{M-1}}\right)^{-1}-2
\right]\dfrac{\mathcal{L}_d N_0}{\Omega \Upsilon_{c}^{CC}} \dfrac{N}{R_{c}^{CC}\log_2 M}+\\
\label{energy_totalcoded}&&(\mathcal{P}_{c}-\mathcal{P}_{Amp})\dfrac{MN}{BR_{c}^{CC}
\log_{2}M}+ 1.75 \mathcal{P}_{Sy}T_{tr}.
\end{eqnarray}

\textbf{LT Codes:} LT codes are the first class of Fountain codes
(designed for erasure channels) that are near optimal erasure
correcting codes \cite{LubyFOCS2002}. The traditional schemes for
data transmission across erasure channels use continuous two-way
communication protocols, meaning that if the receiver can not decode
the received packet correctly, asks the transmitter (via a feedback
channel) to send the packet again. This process continues until all
the packets have been decoded successfully. Fountain codes in
general, and LT codes in particular, surpass the above feedback
channel problem by adopting an essentially one-way communication
approach.

\begin{figure}[t]
\centerline{\psfig{figure=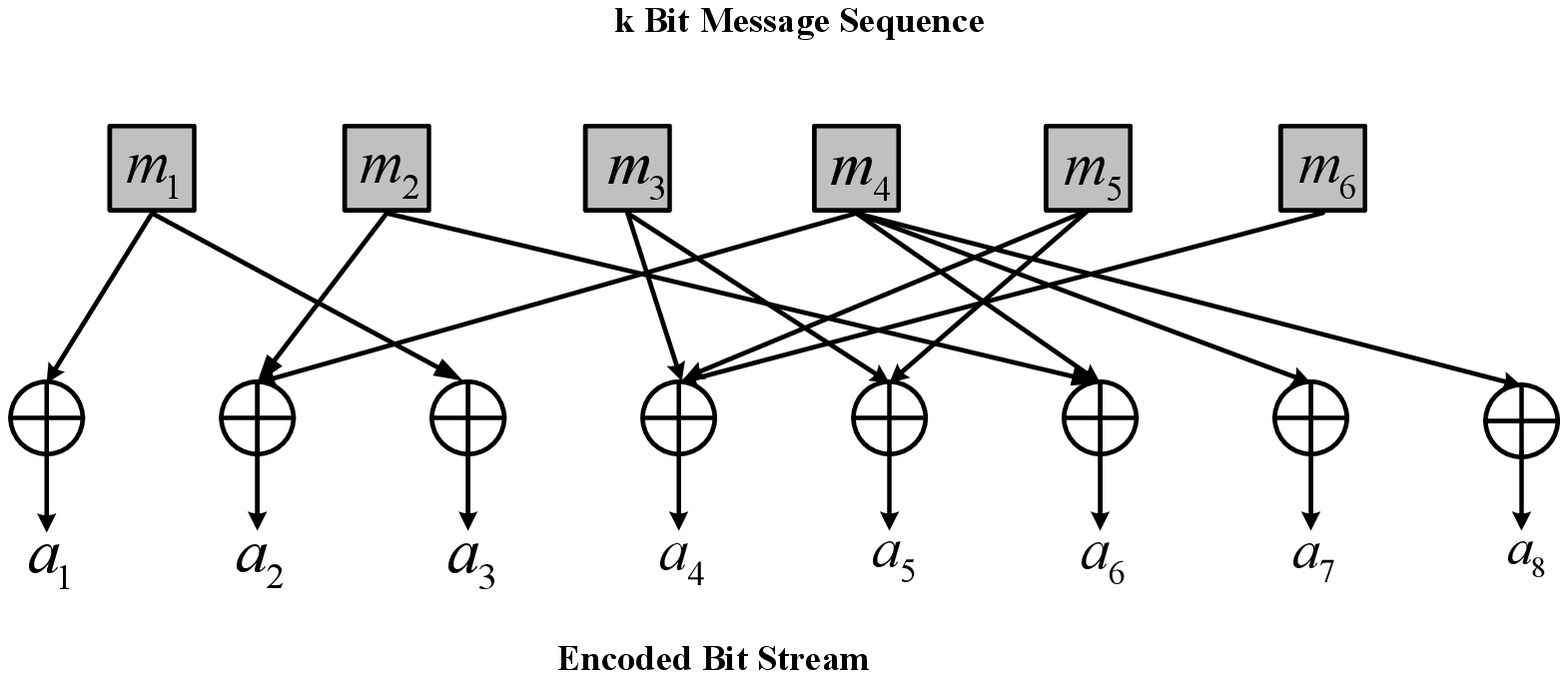,width=4.15in}} \caption{
Bipartite graph of an LT code with $k=6$ and $n=8$. } \label{fig: LT
Encoder}
\end{figure}

LT codes are usually specified jointly by two parameters $k$ (number
of input bits) and $\mathcal{O}(x)$ (the output-node degree
distribution). The encoding process begins by dividing the original
message $\mathcal{M}_N$ into blocks of equal length. Without loss of
generality and for ease of our analysis, we use index $j=1$, meaning
a single $k$-bit message $\mathcal{B}_1 \triangleq
(m_1,m_2,...,m_k)\in \mathcal{M}_N$ is encoded to codeword
$\mathcal{C}_1 \triangleq (a_1,a_2,...,a_n)$. Each single coded bit
$a_i$ is generated based on the encoding protocol proposed in
\cite{LubyFOCS2002}: $i)$ randomly choose a degree $1 \leq
\mathcal{D} \leq k$ from a priori known degree distribution
$\mathcal{O}(x)$, $ii)$ using a uniform distribution, randomly
choose $\mathcal{D}$ distinct input bits, and calculate the encoded
bit $a_i$ as the XOR-sum of these $\mathcal{D}$ bits. The above
encoding process defines a sparse\footnote{If the mean degree
$\mathcal{D}$ is significantly smaller than $k$, then the graph is
sparse.} \emph{bipartite graph} connecting encoded (or equivalently
output) nodes to input nodes (see, e.g., Fig. \ref{fig: LT
Encoder}). It is seen that the LT encoding process is extremely
simple and has very low energy consumption. Unlike classical linear
block and convolutional codes, in which the codeword block length is
fixed, for the above LT code, $n$ is a variable parameter, resulting
in a random variable LT code rate $R^{LT}_c \triangleq \frac{k}{n}$.
More precisely, $a_n \in \mathcal{C}_1$ is the last bit generated at
the output of LT encoder before receiving the acknowledgement signal
from the sink node indicating termination of a successful decoding
process. This inherent property of LT codes means they can vary
their codeword block lengths to adapt to any wireless channel
condition.

We now turn our attention to the LT code design in the proposed WSN.
As seen in the above encoding process, the output-node degree
distribution is the most influential factor in the LT code design.
On the one hand, we need high degree $\mathcal{D}$ in order to
ensure that there are no unconnected input nodes in the graph. On
the other hand, we need low degree $\mathcal{D}$ in order to keep
the number of XOR-sum modules for decoding small. The latter
characteristic means the WSN has less complexity and lower power
consumption. To describe the output-node degree distribution used in
this work, let $\mu_i$, $i=1,...,k$, denote the probability that an
output node has degree $i$. Following the notation of
\cite{Shokrollahi_ITIT0606}, the output-node degree distribution of
an LT code has the polynomial form $\mathcal{O}(x) \triangleq
\sum_{i=1}^{k}\mu_i x^i$ with the property that $\mathcal{O}(1) =
\sum_{i=1}^{k}\mu_i =1$. Typically, optimizing the output-node
degree distribution for a specific wireless channel model is a
crucial task in designing LT codes. The original output-node degree
distribution for LT codes, namely the Robust Soliton distribution
\cite{LubyFOCS2002}, intended for erasure channels, is not optimal
for an error-channel and has poor error correcting capability. In
fact, for wireless fading channels, it is still an open problem,
what the ``optimal" $\mathcal{O}(x)$ is. In this work, we use the
following output-node degree distribution which was optimized for a
BSC using a hard-decision decoder \cite{David_Thesis2008}:
\begin{eqnarray}
\notag \mathcal{O}(x)&=&0.00466x+0.55545x^2+0.09743x^3+0.17506x^5+0.03774x^8+0.08202x^{14}+\\
\label{degree}&&0.01775x^{33}+0.02989x^{100}.
\end{eqnarray}

The LT decoder at the sink node can recover the original $k$-bit
message $\mathcal{B}_1$ with high probability after receiving any
$(1+\epsilon)k$ bits in its buffer, where $\epsilon$ depends upon
the LT code design \cite{Shokrollahi_ITIT0606}. For this recovery
process, the LT decoder needs to correctly reconstruct the bipartite
graph of an LT code. Clearly, this requires perfect synchronization
between encoder and decoder, i.e., the LT decoder would need to know
exactly the randomly generated degree $\mathcal{D}$ for encoding
original $\mathcal{D}$ bits. One practical approach suitable for the
proposed WSN model is that the LT encoder and decoder use identical
pseudo-random generators with a common seed value which may reduce
the complexity further. In this work, we assume that the sink node
recovers $k$-bit message $\mathcal{B}_1$ using a simple
hard-decision ``\emph{ternary message passing}" decoder in a nearly
identical manner to the ``\emph{Algorithm E}'' decoder in
\cite{RichardsonITIT0201} for Low-Density Parity-Check (LDPC)
codes\footnote{Description of the \emph{ternary message passing}
decoding is out of scope of this work, and the reader is referred to
Chapter 4 in \cite{David_Thesis2008} for more details.}. The main
reason for using the ternary decoder here is to make a fair
comparison to the BCH and convolutional codes, since those codes
involved a hard-decision prior to decoding. Also,  the degree
distribution $\mathcal{O}(x)$ in (\ref{degree}) was optimized for a
ternary decoder in a BSC and we are aware of no better
$\mathcal{O}(x)$ for the ternary decoder in Rayleigh fading
channels.

\begin{figure}[t]
\centerline{\psfig{figure=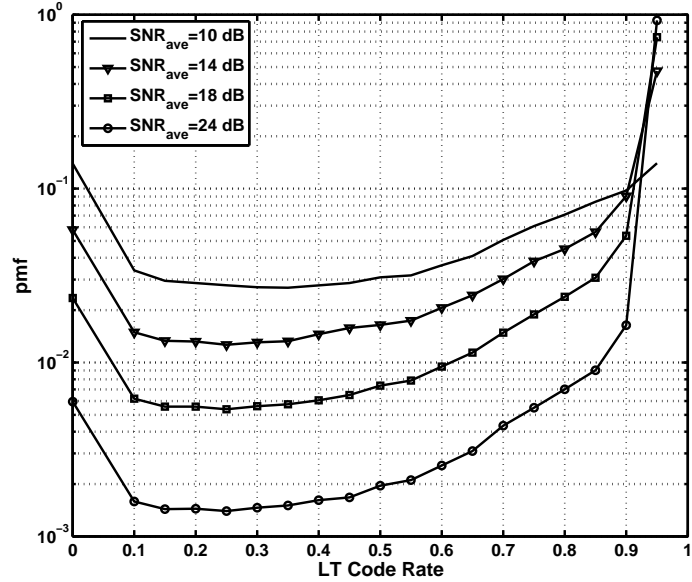,width=4.15in}}
\caption{The pmf of LT code rate for various average SNR and M=2. }
\label{fig: LT code_rate_mpf}
\end{figure}

\begin{figure}[t]
\centerline{\psfig{figure=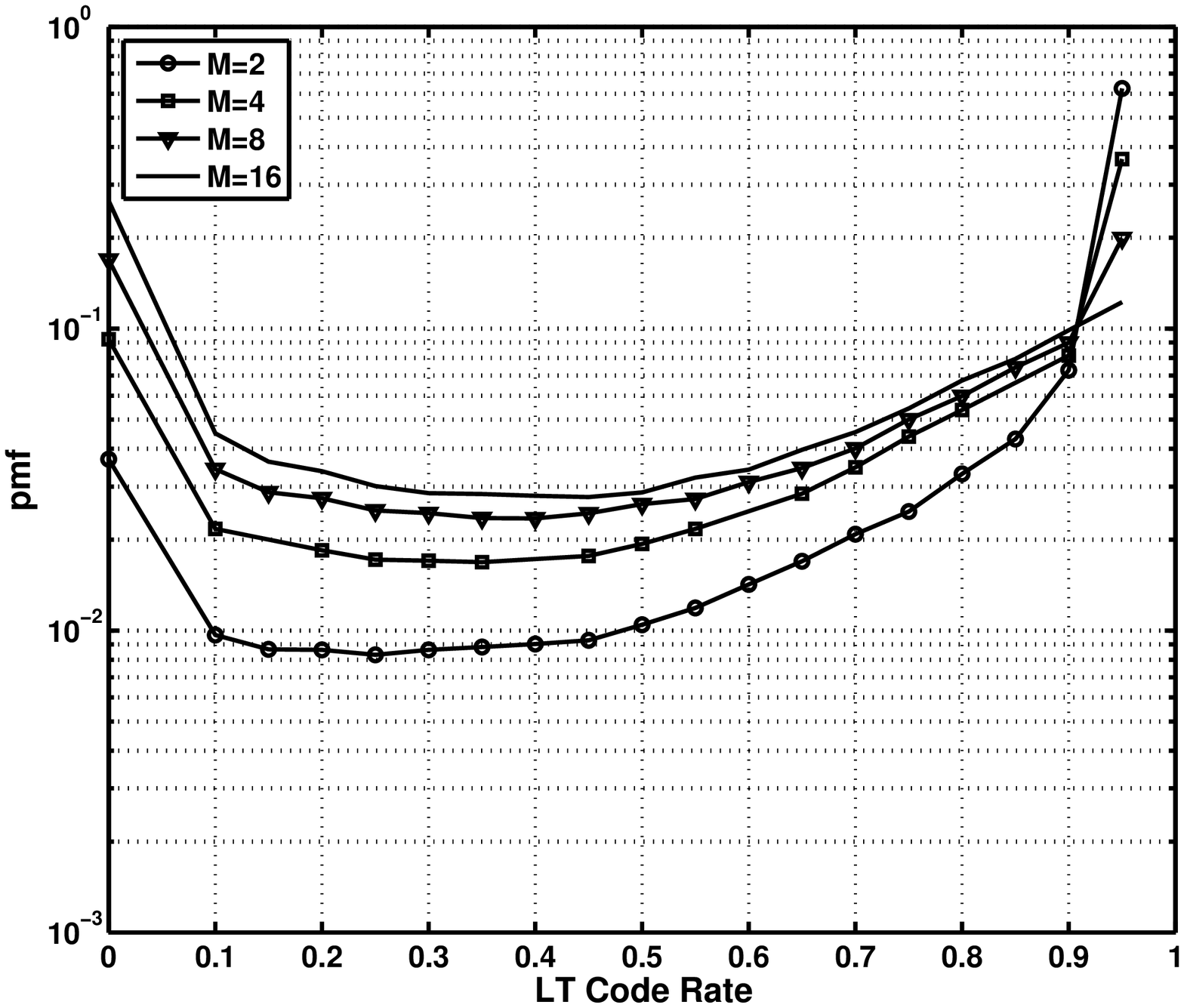,width=4.15in}}
\caption{The pmf of LT code rate for various constellation size $M$
and average SNR=16 dB. } \label{fig: LT code_rate_pmf_M}
\end{figure}

To analyze the total energy consumption of transmitting
$\frac{N}{R_c^{LT}}$ bits during active mode period for LT coded
NC-MFSK scheme, one would compute the LT code rate and the
corresponding LT coding gain. Let us begin with the case of
asymptotic LT code rate, where the number of input bits $k$ goes to
infinity. It is shown that the LT code rate is obtained
asymptotically as a fixed value of $R^{LT}_c \approx
\frac{\mathcal{O}_{ave}}{\mathcal{I}_{ave}}$ for large values of
$k$, where $\mathcal{O}_{ave}$ and $\mathcal{I}_{ave}$ represent the
average degree of the output and the input nodes in the bipartite
graph, respectively (see Appendix II for the proof). When using
finite-$k$ LT codes in a fading channel, the instantaneous SNR
changes from one codeword to the next. Consequently, the rate of the
LT code for any block can be chosen to achieve the desired
performance for that block (i.e., we can collect a sufficient number
of bits to achieve a desired coded BER). This means that for any
given average SNR, the LT code rate is described by either a
probability mass function (pmf) or a probability density function
(pdf) denoted by $P_R(\ell)$. Because it is difficult to get a
closed-form expression of $P_R(\ell)$, we use a discretized
numerical method to calculate the pmf $P_R(\ell)\triangleq
\textrm{Pr}\{R_c^{LT}=\ell \}$, $0 \leq \ell \leq 1$, for different
values of $M$. Table III presents the pmfs of LT code rate for $M=2$
and various average SNR over a Rayleigh fading channel model. To
gain more insight into these results, we plot the pmfs of the LT
code rate in Fig. \ref{fig: LT code_rate_mpf} for the case of $M=2$.
It is observed that for lower average SNRs the pmfs are larger in
the lower rate regimes (i.e., the pmfs spend more time in the low
rate region). Also, all pmfs exhibit quite a spike for the highest
rate, which makes sense since once the instantaneous SNR hits a
certain critical value, the codes will always decode with a high
rate. Also, Fig. \ref{fig: LT code_rate_pmf_M} illustrates the pmf
of LT code rates for various constellation size $M$ and for average
SNR equal to 16 dB. It can be seen that as $M$ increases, the rate
of the LT code tends to have a pmf with larger values in the lower
rate regions.

\begin{table}
\label{table05} \caption{Probability mass function of LT code rate
over Rayleigh fading channel for various average SNR and $M=2$.}
\centering
  \begin{tabular}{|c|c|c|c|c|c|c|c|c|c|c|c|}
  \hline
  Code   &  pmf  &  pmf  &  pmf  &  pmf  &  pmf  &  pmf  &  pmf  &  pmf  &  pmf  &  pmf  &  pmf    \\
  Rate   & (6 dB)& (8 dB)&(10 dB)&(12 dB)&(14 dB)&(16 dB)&(18 dB)&(20 dB)&(22 dB)&(24 dB)&(26 dB)  \\
  \hline
   0.95  & 0.0101 & 0.0521 & 0.1380 & 0.3067 & 0.4742 & 0.6244 & 0.7429 & 0.8290 & 0.8884 & 0.9281 & 0.9540 \\
   0.90  & 0.0180 & 0.0514 & 0.0975 & 0.0978 & 0.0906 & 0.0728 & 0.0536 & 0.0372 & 0.0249 & 0.0164 & 0.0106 \\
   0.85  & 0.0222 & 0.0472 & 0.0841 & 0.0656 & 0.0563 & 0.0431 & 0.0307 & 0.0209 & 0.0138 & 0.0090 & 0.0058 \\
   0.80  & 0.0310 & 0.0540 & 0.0718 & 0.0613 & 0.0499 & 0.0370 & 0.0239 & 0.0174 & 0.0114 & 0.0070 & 0.0047 \\
   0.75  & 0.0442 & 0.0649 & 0.0608 & 0.0617 & 0.0382 & 0.0348 & 0.0189 & 0.0159 & 0.0104 & 0.0054 & 0.0043 \\
   0.70  & 0.0438 & 0.0563 & 0.0506 & 0.0466 & 0.0302 & 0.0248 & 0.0148 & 0.0111 & 0.0072 & 0.0043 & 0.0029 \\
   0.65  & 0.0375 & 0.0439 & 0.0409 & 0.0331 & 0.0244 & 0.0169 & 0.0114 & 0.0075 & 0.0048 & 0.0031 & 0.0020 \\
   0.60  & 0.0370 & 0.0405 & 0.0362 & 0.0285 & 0.0206 & 0.0142 & 0.0095 & 0.0062 & 0.0040 & 0.0026 & 0.0016 \\
   0.55  & 0.0355 & 0.0367 & 0.0317 & 0.0243 & 0.0174 & 0.0119 & 0.0079 & 0.0051 & 0.0033 & 0.0021 & 0.0013 \\
   0.50  & 0.0373 & 0.0369 & 0.0309 & 0.0233 & 0.0164 & 0.0111 & 0.0073 & 0.0048 & 0.0030 & 0.0019 & 0.0012 \\
   0.45  & 0.0345 & 0.0327 & 0.0286 & 0.0197 & 0.0158 & 0.0093 & 0.0065 & 0.0039 & 0.0025 & 0.0016 & 0.0010 \\
   0.40  & 0.0397 & 0.0362 & 0.0278 & 0.0210 & 0.0145 & 0.0097 & 0.0061 & 0.0041 & 0.0026 & 0.0016 & 0.0010 \\
   0.35  & 0.0395 & 0.0346 & 0.0269 & 0.0193 & 0.0132 & 0.0088 & 0.0057 & 0.0037 & 0.0023 & 0.0015 & 0.0009 \\
   0.30  & 0.0421 & 0.0356 & 0.0270 & 0.0192 & 0.0130 & 0.0086 & 0.0056 & 0.0036 & 0.0023 & 0.0015 & 0.0009 \\
   0.25  & 0.0440 & 0.0360 & 0.0278 & 0.0187 & 0.0126 & 0.0083 & 0.0054 & 0.0034 & 0.0022 & 0.0014 & 0.0009 \\
   0.20  & 0.0496 & 0.0393 & 0.0286 & 0.0197 & 0.0132 & 0.0086 & 0.0056 & 0.0035 & 0.0023 & 0.0014 & 0.0009 \\
   0.15  & 0.0541 & 0.0414 & 0.0295 & 0.0201 & 0.0133 & 0.0087 & 0.0056 & 0.0035 & 0.0023 & 0.0014 & 0.0009 \\
   0.10  & 0.0658 & 0.0486 & 0.0338 & 0.0227 & 0.0149 & 0.0097 & 0.0062 & 0.0039 & 0.0025 & 0.0016 & 0.0010 \\
   0.00  & 0.3138 & 0.2115 & 0.1392 & 0.0903 & 0.0579 & 0.0369 & 0.0235 & 0.0149 & 0.0094 & 0.0059 & 0.0038 \\
   \hline
   \end{tabular}
\end{table}

Table IV illustrates the average LT code rates and the corresponding
coding gains of LT coded NC-MFSK using $\mathcal{O}(x)$ in
(\ref{degree}), for $M=2,4,8,16$ and given $P_b=10^{-3}$. The
average rate for a certain average SNR is obtained by integrating
the pmf over the rates from $0$ to $1$. It is observed that the LT
code is able to provide a huge coding gain $\Upsilon_{c}^{LT}$ given
$P_b=10^{-3}$, but this gain comes at the expense of a very low
average code rate, which means many additional code bits need to be
sent. This results in higher energy consumption per information bit.
An interesting point extracted from Table IV is the flexibility of
the LT code to adjust its rate (and its corresponding coding gain)
to suit instantaneous channel conditions in WSNs. For instance in
the case of favorable channel conditions, the LT coded NC-MFSK is
able to achieve $R_{c}^{LT}\approx1$ with $\Upsilon_{c}^{LT}\approx
0$ dB, which is similar to the case of uncoded NC-MFSK, i.e., $n=k$.
In addition, by comparing the results in Table IV with those in
Table II for BCH and convolutional codes, one observes that LT codes
outperform the other coding schemes in energy saving at comparable
rates. The effect of LT code rate flexibility on the total energy
consumption is also observed in the simulation results in the
subsequent section.

\begin{table}
\label{table0021} \caption{Average LT code rate and Coding gain of
LT coded NC-MFSK over Rayleigh flat-fading model for $P_b=10^{-3}$
and M=2,4,8,16.} \centering
  \begin{tabular}{|c|cc||c|cc|cc|cc|}
  \hline
    & M=2 &   & &  M=4 &  & M=8 &  & M=16 & \\
   \hline
   Average  & Average   &  Coding    & Average  & Average   & Coding    & Average   &  Coding    & Average    & Coding    \\
   SNR (dB) & Code Rate &  Gain (dB) & SNR (dB) & Code Rate & Gain (dB) & Code Rate &  Gain (dB) &  Code Rate & Gain (dB) \\
   \hline
   5 & 0.2560 & 25 &  0   & 0.0028 & 33.87 &  0.0012 & 36.46 &  0.0012 & 38.48  \\
   6 & 0.3174 & 24 &  2   & 0.0140 & 31.87 &  0.0024 & 34.46 &  0.0012 & 36.48  \\
   7 & 0.3819 & 23 &  4   & 0.0460 & 29.87 &  0.0095 & 32.46 &  0.0021 & 34.48  \\
   8 & 0.4475 & 22 &  6   & 0.1100 & 27.87 &  0.0330 & 30.46 &  0.0086 & 32.48  \\
   9 & 0.5120 & 21 &  8   & 0.2100 & 25.87 &  0.0870 & 28.46 &  0.0320 & 30.48  \\
   10& 0.5738 & 20 & 10   & 0.3300 & 23.87 &  0.1800 & 26.46 &  0.0870 & 28.48  \\
   11& 0.6315 & 19 & 12   & 0.4600 & 21.87 &  0.3000 & 24.46 &  0.1800 & 26.48  \\
   12& 0.6840 & 18 & 14   & 0.5900 & 19.87 &  0.4400 & 22.46 &  0.3100 & 24.48  \\
   13& 0.7307 & 17 & 16   & 0.7000 & 17.87 &  0.5700 & 20.46 &  0.4500 & 22.48  \\
   14& 0.7716 & 16 & 18   & 0.7800 & 15.87 &  0.6800 & 18.46 &  0.5800 & 20.48  \\
   15& 0.8067 & 15 & 20   & 0.8500 & 13.87 &  0.7700 & 16.46 &  0.6900 & 18.48  \\
   16& 0.8365 & 14 & 22   & 0.8900 & 11.87 &  0.8400 & 14.46 &  0.7800 & 16.48  \\
   17& 0.8614 & 13 & 24   & 0.9200 & 9.87  &  0.8800 & 12.46 &  0.8400 & 14.48  \\
   18& 0.8821 & 12 & 26   & 0.9400 & 7.87  &  0.9100 & 10.46 &  0.8800 & 12.48  \\
   19& 0.8991 & 11 & 28   & 0.9500 & 5.87  &  0.9300 & 8.46  &  0.9200 & 10.48  \\
   20& 0.9130 & 10 & 30   & 0.9600 & 3.87  &  0.9500 & 6.46  &  0.9400 & 8.48   \\
   22& 0.9333 & 8  & 32   & 0.9600 & 1.87  &  0.9500 & 4.46  &  0.9500 & 6.48   \\
   24& 0.9466 & 6  & 34   & 0.9600 & -0.13 &  0.9600 & 2.46  &  0.9500 & 4.48   \\
   26& 0.9551 & 4  & 36   & 0.9700 & -2.13 &  0.9600 & 0.46  &  0.9600 & 2.48   \\
   28& 0.9606 & 2  & 38   & 0.9700 & -4.13 &  0.9700 & -1.54 &  0.9600 & 0.48   \\
   30& 0.9640 & 0  & 40   & 0.9700 & -6.13 &  0.9700 & -3.54 &  0.9700 & -1.52  \\

   \hline
   \end{tabular}
\end{table}

Unlike BCH and convolution codes in which the active mode duration
of coded NC-MFSK is fixed, for the LT coded NC-MFSK, we have
non-fixed values for $T_{ac}^{LT}=\frac{N}{b R_{c}^{LT}}T_s$. With a
similar argument as for BCH and convolutional codes, the total
energy consumption of transmitting $\frac{N}{R_{c}^{LT}}$ bits for a
given $P_b$ is obtained as a function of the random variable
$R_{c}^{LT}$ as follows:
\begin{eqnarray}
\notag \mathcal{E}^{LT}_N &=& (1+\alpha) \left[\left(
1-\left(1-\dfrac{2(M-1)}{M}P_b\right)^{\frac{1}{M-1}}\right)^{-1}-2
\right]\dfrac{\mathcal{L}_d N_0}{\Omega \Upsilon_{c}^{LT}} \dfrac{N}{R_{c}^{LT}\log_2 M}+\\
\label{energy_totalcoded1}&&(\mathcal{P}_{c}-\mathcal{P}_{Amp})\dfrac{MN}{BR_{c}^{LT}
\log_{2}M}+ 1.75 \mathcal{P}_{Sy}T_{tr},
\end{eqnarray}
where the goal is to minimize the average $\mathcal{E}^{LT}_N$ over
$R_{c}^{LT}$.

\section{Numerical Results}\label{simulation_Ch5}
In this section, we present some numerical evaluations using
realistic parameters from the IEEE 802.15.4 standard and
state-of-the art technology to confirm the energy efficiency
analysis of uncoded and coded NC-MFSK modulation schemes discussed
in Sections \ref{uncoded_MFSK} and \ref{analysis_Ch4}. We assume
that the NC-MFSK modulation scheme operates in the $f_0=$2.4 GHz
Industrial Scientist and Medical (ISM) unlicensed band utilized in
IEEE 802.15.14 for WSNs \cite{IEEE_802_15_4_2006}. According to the
FCC 15.247 RSS-210 standard for United States/Canada, the maximum
allowed antenna gain is 6 dBi \cite{FreeScale2007}. In this work, we
assume that $\mathcal{G}_t=\mathcal{G}_r=5$ dBi. Thus for $f_0=$2.4
GHz, $\mathcal{L}_1~ \textrm{(dB)}\triangleq
10\log_{10}\left(\frac{(4 \pi)^2}{\mathcal{G}_t \mathcal{G}_r
\lambda^2}\right) \approx 30~ \textrm{dB}$, where $\lambda
\triangleq \frac{3\times 10^8}{f_0}=0.125$ meters. We assume that in
each period $T_N$, the sensed data frame size $N=1024$ bytes (or
equivalently $N=8192$ bits) is generated, where $T_N$ is assumed to
be 1.4 seconds. The channel bandwidth is assumed to be $B=62.5$ KHz,
according to IEEE 802.15.4 \cite[p. 49]{IEEE_802_15_4_2006}. From
$\frac{2^{b_{max}}}{b_{max}}=\frac{B}{N}(T_{N}-T_{tr})$, we find
that $M_{max} \approx 64$ (or equivalently $b_{max} \approx 6$) for
NC-MFSK. The power consumption of the LNA and IF amplifier are
considered 9 mw \cite{Bevilacqua_IJSSC1204} and 3 mw
\cite{Cui_GoldsmithITWC0905, TangITWC0407}, respectively. The power
consumption of the frequency synthesizer is supposed to be 10 mw
\cite{Wang_ISLPED2001}. Table V summarizes the system parameters for
simulation. The results in Tables II-IV are also used to compare the
energy efficiency of uncoded and coded NC-MFSK schemes.

\begin{table}
\label{table001} \caption{System Evaluation Parameters} \centering
  \begin{tabular}{lll}
  \hline
   $B=62.5$ KHz          & $N_0=-180$ dB               & $\mathcal{P}_{ADC}=7$ mw \\

   $M_l=40$ dB           & $\mathcal{P}_{Sy}=10$ mw    &  $\mathcal{P}_{LNA}=9$ mw\\

   $\mathcal{L}_1=30$ dB & $\mathcal{P}_{Filt}=2.5$ mw & $\mathcal{P}_{ED}=3$ mw  \\

   $\eta=3.5$            & $\mathcal{P}_{Filr}=2.5$ mw & $\mathcal{P}_{IFA}=3$ mw \\

   $\Omega=1$            &  $T_N=1.4$ sec              & $T_{tr}=5~\mu s$      \\

  \hline
  \end{tabular}
\end{table}

\begin{figure}[t]
\centerline{\psfig{figure=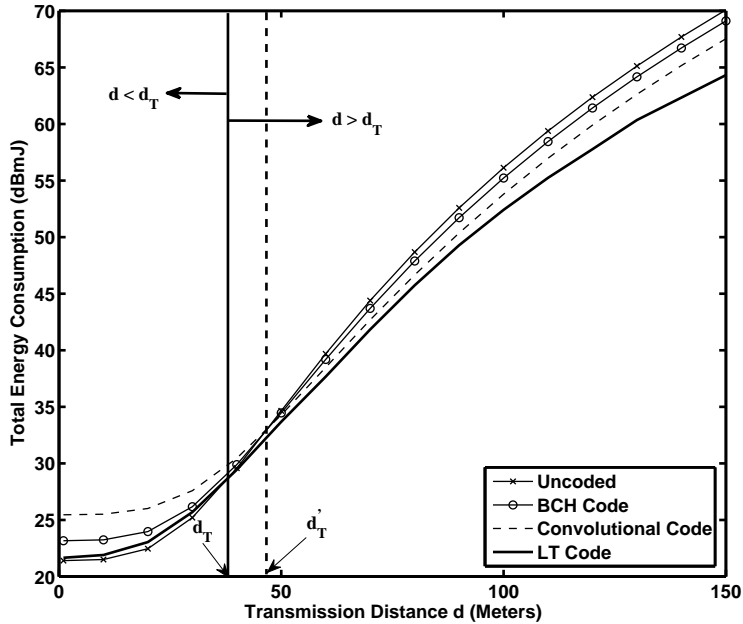,width=4.5in}}
\caption{Total energy consumption of optimized coded and uncoded
NC-MFSK versus $d$ for $P_b=10^{-3}$.} \label{fig: Coded_Uncoded}
\end{figure}

Fig. \ref{fig: Coded_Uncoded} shows the total energy consumption
versus distance $d$ for the optimized BCH, convolutional and LT
coded NC-MFSK schemes, compared to the optimized uncoded NC-MFSK for
$P_b=10^{-3}$. The optimization is done over $M$ and the parameters
of coding scheme. Simulation results show that for $d$ less than the
threshold level $d_T \approx 40$ m, the total energy consumption of
optimized uncoded NC-MFSK is less than that of the coded NC-MFSK
schemes. However, the energy gap between LT coded and uncoded
NC-MFSK is negligible compared to the other coded schemes as
expected. For $d>d_T$, the LT coded NC-MFSK scheme is more
energy-efficient than uncoded and other coded NC-MFSK schemes. Also,
it is observed that the energy gap between LT and convolutional
coded NC-MFSK increases when the distance $d$ grows. This result
comes from the high coding gain capability of LT codes which
confirms our analysis in Section \ref{analysis_Ch4}. The threshold
level $d_T$ (for LT code) or $d^{'}_T$ (for BCH and convolutional
codes) are obtained when the total energy consumptions of coded and
uncoded systems become equal. For instance, using $\mathcal{L}_d
=M_ld^\eta \mathcal{L}_1$, and the equality between
(\ref{energy_totFSK}) and (\ref{energy_totalcoded}) for uncoded and
convolutional coded NC-MFSK, we have
\begin{equation}
d^{'}_T=\left[\dfrac{M \Omega
(\mathcal{P}_{c}-\mathcal{P}_{Amp})}{(1+\alpha) \left[\left(
1-\left(1-\dfrac{2(M-1)}{M}P_b\right)^{\frac{1}{M-1}}\right)^{-1}-2
\right]}\dfrac{1}{\mathcal{L}_1
M_l}\dfrac{\Upsilon_{c}^{CC}(1-R_c^{CC})}{\Upsilon_{c}^{CC}R_c^{CC}-1}\right]^{\frac{1}{\eta}}.
\end{equation}

It should be noted that the above threshold level imposes a
constraint on the design of the physical layer of some wireless
sensor networking applications, in particular dynamic WSNs. To
obtain more insight into this issue, let assume that the location of
the sensor node is changed every $T_d \gg T_c$ time unit, where
$T_c$ is the channel coherence time. For the moment, let us assume
that the sensor node aims to choose either a \emph{fixed-rate coded}
or an uncoded NC-MFSK based on the distance between sensor and sink
nodes. According to the results in Fig. \ref{fig: Coded_Uncoded}, it
is revealed that using fixed-rate channel coding is not energy
efficient for short distance transmission (i.e., $d < d^{'}_T$),
while for $d > d^{'}_T$, convolutional coded NC-MFSK is more
energy-efficient than other schemes. For this configuration, the
sensor node must have the capability of an adaptive coding scheme
for each distance $d$. However, as discussed previously, the LT
codes can adjust their rates for each channel condition and have
(with a good approximation) minimum energy consumption for every
distance $d$. This indicates that LT codes can surpass the above
distance constraint for WSN applications with dynamic position
sensor nodes over Rayleigh fading channels. This characteristic of
LT codes results in reducing the complexity of the network design as
well. Of interest is the strong benefits of using LT coded NC-MFSK
compared with the coded modulation schemes in
\cite{Cui_GoldsmithITWC0905, Chouhan_ITWC1009}. In contrast to
classical fixed-rate codes used in \cite{Cui_GoldsmithITWC0905,
Chouhan_ITWC1009}, the LT codes can vary their block lengths to
adapt to any channel condition in each distance $d$. Unlike
\cite{Cui_GoldsmithITWC0905} and \cite{Chouhan_ITWC1009}, where the
authors consider fixed-rate codes over an AWGN channel model, we
considered a Rayleigh fading channel which is a general model in
practical WSNs. The simplicity and flexibility advantages of LT
codes with an NC-MFSK scheme make them the preferable choice for
wireless sensor networks, in particular for WSNs with dynamic
position sensor nodes.

\section{Conclusion}\label{conclusion_Ch6}
In this paper, we analyzed the energy efficiency of LT coded NC-MFSK
in a proactive WSN over Rayleigh fading channels with path-loss. It
was shown that the energy efficiency of LT codes is similar to that
of uncoded NC-MFSK scheme for $d<d_T$, while for $d > d_T$, LT coded
NC-MFSK outperforms other uncoded and coded schemes, from the energy
efficiency point of view. This result follows from the flexibility
of the LT code to adjust its rate and the corresponding LT coding
gain to suit instantaneous channel conditions for any transmission
distance $d$. This rate flexibility offers strong benefits in using
LT codes in practical WSNs with dynamic distance and position
sensors. In such systems and for every value of distance $d$, LT
codes can adjust their rates to achieve a certain BER with low
energy consumption. The importance of our scheme is that it avoids
some of the problems inherent in adaptive coding or Incremental
Redundancy (IR) systems (channel feedback, large buffers, or
multiple decodings), as well as the coding design challenge for
fixed-rate codes used in WSNs with dynamic position sensor nodes.
The simplicity and flexibility advantages of LT codes make the LT
code with NC-MFSK modulation can be considered as a \emph{Green
Modulation/Coding} (GMC) scheme in dynamic WSNs.

\appendices

\section{Energy Consumption of BCH Decoder}\label{decoder}
For the sink circuitry, we have an extra energy cost due to the
decoding process in a coded NC-MFSK scheme. It is shown in \cite[p.
160]{Karl_Book2005} that the energy consumption of an BCH$(n,k,t)$
decoder per codeword, denoted by $\tilde{\mathcal{E}}_{Dec}^{BC}$,
is computed as $\tilde{\mathcal{E}}_{Dec}^{BC}=\left(2nt+2t^2
\right)\left(\mathcal{E}_{add}+\mathcal{E}_{mult} \right)$, where
$\mathcal{E}_{add}$ and $\mathcal{E}_{mult}$ represent the energy
consumptions of adder and multiplier in unit of W/MHz, respectively.
For instance, the energy consumption of 0.5 $\mu$W/MHz indicates
that the consumed energy per clock cycle is 0.5 pJ. Thus, if for
decoding of each bit we consider 20 clock cycles, the energy
consumption is 10 pJ/bit. For this case, the total energy
consumption a BCH$(n,k,t)$ decoder to recover $N$-bit message
$\mathcal{M}_N$ is obtained as
$\mathcal{E}_{Dec}^{BC}=\frac{N}{k}\tilde{\mathcal{E}}_{Dec}^{BC}=\frac{N}{k}\left(2nt+2t^2
\right)\left(\mathcal{E}_{add}+\mathcal{E}_{mult} \right)$. It is
shown in \cite{Meir_CICC1996} that the energy consumption per
addition or multiplication operation is on the order of pJ per bit.
According to the values of $n$ and $t$ in Table II, one can with a
good approximation assume that the energy consumption of an BCH
decoder is negligible compared to the other circuit components in
the sink node.

\section{Asymptotic LT Code Rate}
The proof of the remark is straightforward using the notation of
\cite{Etesami_ITIT0506} and the bipartite graph concepts in graph
theory. Obviously, the output-node degree distribution
$\mathcal{O}(x)$ induces a distribution on the input nodes in the
bipartite graph. Thus, in the asymptotic case of $k \to \infty$, we
have the input-node degree distribution defined as $\mathcal{I}(x)
\triangleq \sum_{i=1}^{\infty}\nu_i x^i$, where $\nu_i$ denotes the
probability that an input node has a degree $i$. In this case, the
average degree of the input and output nodes are computed as
$\frac{d \mathcal{I}(x)}{d x}\big|_{x=1} \triangleq
\mathcal{I}_{ave}$ and $\frac{d \mathcal{O}(x)}{d x}\big|_{x=1}
\triangleq \mathcal{O}_{ave}$, respectively. Thus, the number of
edges exiting the input nodes of the bipartite graph, in the
asymptotic case of $k \to \infty$, is $k\mathcal{I}_{ave}$, which
must be equal to $n\mathcal{O}_{ave}$, the number of edges entering
the output nodes in the graph. As a results, the asymptotic LT code
rate is obtained as $R^{LT}_c= \frac{k}{n} \approx
\frac{\mathcal{O}_{ave}}{\mathcal{I}_{ave}}$, which is a
deterministic value for given $\mathcal{O}(x)$ and $\mathcal{I}(x)$.

%\bibliographystyle{IEEE}
%\bibliography{keylatex}

\end{document}